\documentstyle[aps,prl,tighten,psfig]{revtex}
\begin{document}

\title{
Depolarisation of
spherical--membrane quantum well:
Gap renormalisation for closed--shell fullerenes
}

\author{Slava V. Rotkin}

\address{Ioffe Institute, 26 Polytehnicheskaya, St.Petersburg
194021, Russia;\\
Beckman Institute, UIUC, 405 N.Mathews, Urbana, IL 61801, USA. \\
{\ E-mail: \ \ rotkin@uiuc.edu}
}

\date{December 31, 1999}

\maketitle

\begin{abstract}
An anomalous large level shift is newly found to correct
the one--electron spectrum of a closed--shell carbon nanocluster,
for example, the spectrum which was described within the model of
a thin quantum well rolled into a sphere. As a result of the
interaction with zero--point oscillations of the confined modes
of the electric field, the one--electron energy levels heighten.
The depolarization depends on the electron momentum, why it does
not shift the rotational spectrum as a whole.  Surprisingly the
shift in a spherical closed--shell cluster of an arbitrary size
is described by an universal law.  The non-equidistant shift of
the levels results in an increase of an one-electron gap by 1.4
times.
\end{abstract}

%
%

\section{Introduction}
\label{sec:intro}

One--electron approximation (owing to its cheapness)
is often used for carbon
nanoclusters though
at an expense of uncontrolled incorrectness. It is
worth to be aware when the simple approach fails because a true
many--body calculation seems to be too complicated to apply it
for any new cluster appearing in the large fullerene
family. While a complete account for Coulomb interaction has to
include all renormalization effects, a summing of some of
diagrams can lead to the depolarization lost. A model estimation
will be performed below which captures some physics, usually
covered only by the sophisticated many--body theory. We stress
that our term is a counterpart of the standard vertex
renormalization ({\it i.e.}, electron--hole attraction), as
increasing the one--electron gap.

In the paper we
go to reveal the depolarization correction which follows from an
interaction of an electron with an electromagnetic field created
by all other valence electrons of the closed--shell cluster.
Therefore, the electron interaction is treated selfconsistently
within the approach. This continues our consideration of C$_{60}$
in frame of a simple quantum mechanical model, namely, the model
of the spherical--membrane quantum well (SMQW) \cite{rotkftt},
which has fruitful analogies with a standard quantum well model
in the theory of low--dimensional structures.

The group of full rotations, SO(3), was shown to be useful
in the fullerene physics to label the one--electron
states\cite{rotkftt,dress,bertch}, to simplify a theory of an
electron--electron interaction on the sphere\cite{assa} and
surface plasmons\cite{barton}, as well as to facilitate a
computation of a high--harmonic generation spectrum in the
fullerene\cite{rkfp}. An essential simplification is achieved
(also in the depolarization calculation) using the spherical
symmetry because of the angular momentum subspaces\cite{varshal}
can be separated readily in many cases.  For a linear dipole
response, the SO(3) symmetry cancels a lot of matrix elements
and allows one to get analytically the solution for the
selfconsistent RPA response function of
C$_{60}$\cite{rotkftt,bertch,bulgak,molmat}. A peak of a
collective excitation shows up in this spectrum, resulting from
fast coherent oscillations of a total electron density of valence
states.

This surface density oscillation can be thought as a confined
electrical field mode or the surface plasmon (which was studied
before\cite{rotkftt,bertch,barton,plasm}). A zero--point
oscillation of an electromagnetic vacuum is well known to
manifest itself as a Casimir force between close surfaces of a
polarizable substance, as a van--der--Waals interaction, as a
standard Lamb shift in a hydrogen--like atom. We will consider in
the paper the shift of electron levels in the field
%
%
fluctuations of the confined modes (connected with
the nanocluster), which effect is much stronger than of the
zero--point oscillations of the free electromagnetic field.

Below we will show that, within the spherical--symmetry of SMQW
model, the one--electron level shift due to the interaction with
the zero--point oscillations of the electric field of the local
collective mode (or the depolarization) results in a strong
renormalization of a gap of the closed--shell fullerene. The
relative gap increment is independent of the cluster size.

\section{Perturbation theory for energy level shift}
\label{sec:ls}

\begin{figure}[htb]
\centerline{ \psfig{figure=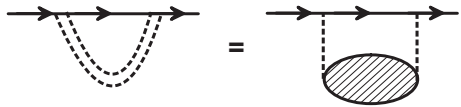,width=6cm}}
\vskip .1cm
\caption{\label{fig:pla}
The diagram representing the depolarization level shift of the
SMQW electron. Within our approximation
the shift results from the electric field
of the surface plasmon mode
which is depicted as the shaded mass operator in the right. Due
to the imposed spherical symmetry the angular momentum conserves
for the two--leg diagram.} \end{figure}

We put forward a semiclassical theory of an energy level shift
(LS) for an arbitrary shell object in Ref.\cite{hawaii}, keeping
the simplicity of the one--electron calculation and outlining
the depolarization. The method follows the book\cite{migdal}. For
completeness we give here the bases of the computation.
The frequency of the (zero--point) oscillations
of the external field is much higher
than the inverse period of the electron orbit.  Therefore, the
adiabatic approximation has to be used. It means that the fast
(field) variables can be integrated out in a motion equation for
the (slow) electron. A simple model for the depolarization
states that the LS results from short fast deflections of the
electron from its original orbit in the random high--frequency
field of the electromagnetic wave.
The energy correction is given by the second order perturbation
theory (see the diagram in Fig.\ref{fig:pla}) and reads as:
\begin{equation}
\overline{\Delta H} = \left< H(r+\delta)-H(r) \right> =
\left< \nabla H\cdot \vec\delta + \frac{1}{2} \nabla^2 H
\; \vec \delta \cdot \vec \delta + \ldots\right> =
\frac{1}{4} \nabla^2 H \; \overline{\delta^2}
+o(\overline{\delta^2})
\label{ls1}
\end{equation}
where $H(r)$ is the unperturbed (one--electron) Hamiltonian and
$H(r+\delta)$ is the Hamiltonian with account for the electron
deflection $\delta$. The energy difference is expanded in
series on $\delta$, then averaged over the fluctuations and a
first nonzero contribution is taken. The small parameter of the
perturbation theory will be proved in the end of the section. It
is the ratio of the deflection to a characteristic length of the
potential which estimates Laplasian, $\overline{\delta^2}\nabla^2
\sim \overline{\delta^2}/R^2_C \ll 1$.
We stress that the spherical
symmetry allows us to limit the calculation to the subspace of
the fixed angular momentum as well as to get the eigen modes of
the confined field. The angular momentum plays the role of the
simple momentum for a space invariant system. It conserves for
definite types of diagrams (for example, the bubble diagrams).
That means that the spherical plasmon modes do not mix.

The expression for the mean square of the deflection,
$\overline{\delta^2}$, caused by the zero--point fluctuation of
the confined electromagnetic modes, was deduced in
Ref.\cite{hawaii}.  A contribution from the free electromagnetic
field, which will not be considered below, was also calculated
and shown to be miserable comparing with the confined field.  We
will not repeat the derivation completely but refer to the
expression (5) of that article, which gives the deflection from
the spherical modes, $|L\rangle$, as follows:
\begin{equation}
\overline{\delta^2}=\sum \limits_{L=1}^{L_c} (2L+1)
\overline{\delta^2_L}
\simeq \displaystyle \frac{\pi 2^{3/2}}{3}
\left( \frac{R}{N a_B}\right)^{3/2}
a_B^2 \;
\left(L_{\rm c}+\frac{1}{2}\right)^{3/2},
\label{delta}
\end{equation}
where $R$ is a nanocluster radius, $a_B=\hbar^2/m e^2$ is the
Bohr radius (atomic length unit), $L_{\rm c}$ is the maximum
allowed angular momentum of the plasmon state. A number of
atoms of the cluster, $N$, reads as follows:
\begin{equation}
N=\frac{4\pi R^2}{3\sqrt{3} \; b^2/4},
\label{numb}
\end{equation}
where $b\sim 1.4$~\AA~ is the carbon--carbon distance in the
graphite--like lattice of the nanocluster. Using this definition
we are able to evaluate the mean square deflection. For the
infinitely large cluster ($R,N\to\infty$) the (infinitely large)
angular momentum can be related to the (finite) 2D wave--number
of the surface excitation via: ${\hat L}\simeq {\hat k} R$.
The maximum wave--number $k_{\rm max}\sim
\pi/\sqrt{3}b$
lies on the "Brilluene zone"\cite{bz}
boundary.
Substituting
the corresponding angular momentum value $L_{\rm c}
+1/2\sim \pi R/(b\sqrt{3})$ and the number of atoms
into the expression (\ref{delta}) we get the semiclassical
value of the electron deflection. In the units of the atomic
length it reads as:
\begin{equation}
\displaystyle \frac{\overline{\delta^2}}{a_B^2}
= \displaystyle \frac{3^{5/4}}{2^{9/2}\sqrt{\pi}}
\left(\frac{b}{a_B}\right)^{3/2}
\left(\frac{b}{R}\right)^{3/2}
\left(L_{\rm c}+\frac{1}{2}\right)^{3/2}
=\displaystyle \frac{3^{1/2}\pi}{2^{9/2}}
\left(\frac{b}{a_B}\right)^{3/2}\simeq 1.03.
\label{delta2}
\end{equation}
Though the estimation is semiquantitative, the deflection seems
to be of the order of the atomic unit, which proves the expansion
(\ref{ls1}). It follows from Eq.(\ref{delta2}) that the
perturbation theory works as long as the potential changes on the
scale larger than the atomic one.

The first result of the model is that the mean square
deflection of the electron in the SMQW does not depend on the
radius, neither on the number of atoms. It is ocularly because of
the density of the valence electrons is constant (precisely, it
grows slightly with $N$ reflecting the fact that the hexagonal
carbon lattice of the spherical cluster includes 12 pentagons
those lessen the density, which becomes insignificant for the
large enough cluster). The independence of the deflection on the
number of atoms follows from the extreme quantization both of the
electron and the field mode.

\section{Level ordering in SMQW and angular momentum
dependent shift}
\label{sec:shift}

Suppose that the one--electron
model works for some cluster C$_{N}$.
To make a numerical estimation we will think about C$_{60}$,
which spectrum was well studied experimentally. The result does
not depend essentially on the
one--electron model chosen, therefore,
in order to present
a manifestation of the depolarization, the simplest SMQW
model will be used for the bare level ordering.
Then the one--electron Hamiltonian
reads as \cite{rotkftt}:
\begin{equation}
H_o=E_n+ \frac{\hbar^2}{2mR^2} \hat L^2,
\label{SMQW}
\end{equation}
where $E_n$ is the energy of a lowest level of $n-$th radial
series; an orbital quantization energy $\hbar \omega_o\equiv
\hbar^2/mR^2$ defines the SO(3) level spacing between states
$|n,LM\rangle$ which are the eigenstates of the angular momentum
operator and are $2L+1$ degenerated. We will refer below to the
single series with $n=1$ corresponding to $\pi$ electron system
of the nanocluster, therefore the radial index will be omitted.
It will be convenient to substitute the classical
value $L+1/2$ for the angular momentum operator
eigenvalues\cite{varshal}, which is correct for the large enough
momentum.

Let us rewrite the orbital energy of the $L-$th electron state in
the following form:
\begin{equation}
E^{(o)}_L=\frac{\hbar \omega_o}{2} \left( L+\frac{1}{2}\right)^2
\simeq \frac{8\pi}{3\sqrt{3}} \left( \frac{a_B}{b}\right)^2
\frac{\left( L+\frac{1}{2}\right)^2}{N} E_B=
E^{(o)}_{\rm max}
\frac{\left( L+\frac{1}{2}\right)^2}{N},
\label{so3level}
\end{equation}
%
%
where $E_B=e^2/a_B$ is the atomic energy unit. The meaning of the
energy
$E^{(o)}_{\rm max}=E_B (a_B/b)^2 8\pi/3\sqrt{3}\simeq 16.8$~eV
will become clear in the end of the section.
The expression (\ref{so3level}) is
derived using the surface carbon density
appearing in the denominator of Eq.(\ref{numb}).

The electrons move within a very thin spherical shell layer
(spherical membrane) which is approximated by a delta--function
$\delta(r-R)$. Hence, we use 2D--Laplasian
operator in Eq.(\ref{ls1}),
which is nothing more than its angular part
in the radial co--ordinate system $\hat L^2/R^2$.
Evidently, one has $\langle \nabla^2 H_o\rangle\sim
\frac{\hbar^2}{2mR^2}\frac{1}{R^2} \left(L+\frac{1}{2}\right)^4$.
Finally, Eq.(\ref{ls1}) yields the SMQW level shift as
follows:
\begin{equation}
\displaystyle \delta E_L = \frac{\pi^3
\sqrt{2}}{9\sqrt{3}} \left( \frac{a_B}{b}\right)^{5/2}
\frac{\left( L+\frac{1}{2}\right)^4}{N^2} E_B,
\label{SMQWls}
\end{equation}
here the last fraction is dimensionless. As it will be
shown, it is actually independent of the nanocluster size for
some characteristic $L$, {\it e.g.} for $L_F$ --- Fermi momentum
dividing the empty levels from the occupied ones.

The second result of the model is that the universal law for the
level shift is independent on the specific nanocluster (excluding
the trivial dependence in $N$ which also drops as will
be explained
in the end of the section). The depolarisation
heightens the one--electron energy as:
\begin{equation}
E_L= E_L^{(o)} \left( 1+\kappa \frac{\hat L^2}{N} \right),
\label{unilaw}
\end{equation}
where $\kappa\sim 0.36$ is the numerical coefficient depending
only on the carbon atom density: $\kappa =\sqrt{a_B/b}\;
\pi^2/(2^{2.5}3)$. In the contraction limit $L\to kR$, the rest
term goes to the squared wave--number $\hat L^2/N \to (kb)^2$
with the accuracy of some factor.
%
%
%
The lowest level does not shift because of the zero
value of the angular momentum (in contrast to the standard Lamb
shift which originates from the interaction with the charge of
the hydrogen--like nuclei core. This charge is concentrated in
the co--ordinate origin, therefore, the maximum Lamb shift is for
the lowest, $s$, state). Note that the $L=0$ term is absent
in Eq.(\ref{delta}) owing to no monopole plasmon exists.

\begin{figure}[htb]
\centerline{ \psfig{figure=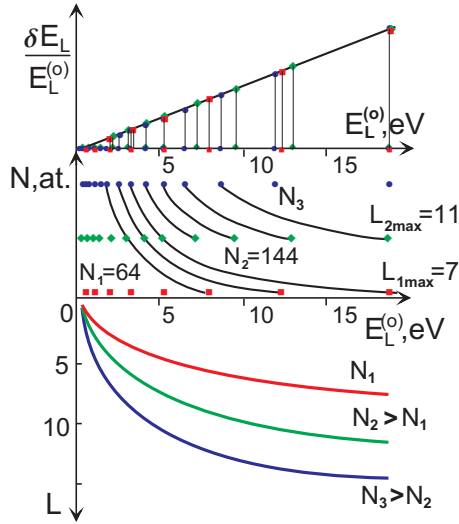,width=6cm}}
\vskip .1cm
\caption{\label{fig:parabola}
The universal law for the depolarization level shift. Bottom: The
bare SMQW electron energy levels for 3 different cluster sizes:
$N_1=64, N_2=144, N_3=196$. The number of atoms is chosen to form
the closed shells. Note that the upper energy $E_{\rm max}^{(o)}$
is the same for all clusters. Middle: Fan--like diagram for the
energy levels of the fixed state $|L\rangle$, depending on $N$.
Top: The level shift depends on the cluster size and $L$ but all
data falls into the same straight line.} \end{figure}

Let us evaluate the maximum depolarization LS.
It occurs for the maximum
angular momentum $L_{\rm max}+1/2$ which is derived from the sum
of the electron states. Because the total number of
(double degenerate due to the spin) $\pi-$states
equals the number of carbon atoms,
$N= \sum\limits_{L=0}^{L_{\rm max}} (2L+1)= (L_{\rm max}+1)^2$,
we substitute for
$L_{\rm max}+1/2\sim \sqrt{N}$.
Now the meaning of the energy $E_{\rm max}^{(o)}$, entered
Eq.(\ref{so3level}), is clear. It is the upper limit for the bare
energy (see spectra in Fig.\ref{fig:parabola}).
From Eq.(\ref{unilaw}) the maximum depolarization LS
reads as:
\begin{equation}
\displaystyle
\frac{\delta E_L}{E_L^{(o)}} \le
\frac{\pi^2}{12\sqrt{2}} \sqrt{\frac{a_B}{b}}
\frac{(L_{\rm max}+1/2)^2}{N}= \kappa
\simeq 0.36.
\label{maxls}
\end{equation}
As we claimed before, the LS for the fixed (upper) level does
not depend neither on the state label $L$ nor on the cluster size
$N$. Let us now reflect on the size scaling of the shift of the
state $|L\rangle$. The only state with $L$ increasing as
$\sqrt{N}$ has a physical sense because of such momentum, scaling
with the cluster size, remains in the same point of the
"Brilluene zone". The LS of this fixed state constitutes the
fixed percentage (of the bare energy) which is equal for any
cluster size. The depolarisation for the different $L$ varies
from 0 for the lowest state of a closed--shell cluster, to
$\kappa$ for the upper one. The universality for our scaling law
means that for an arbitrary cluster size and an arbitrary state
momentum the relative LS falls into the same straight line as
shown in Fig.\ref{fig:parabola}.
Clearly, Eq.(\ref{maxls}) proves that the perturbation theory is
applicable as its correction is still less than unity for the
highest possible level which has the maximum shift.

\section{Gap increase within SMQW--depolarization model}
\label{sec:SMQWgap}

\begin{figure}[htb]
\centerline{ \psfig{figure=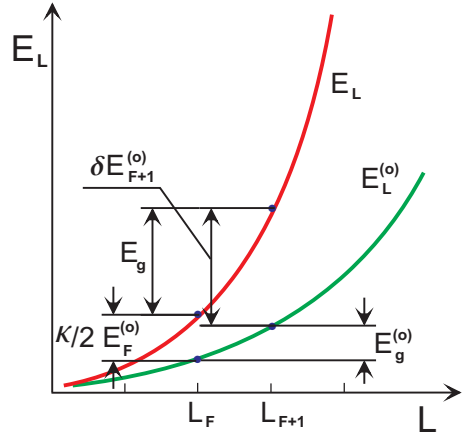,width=6cm}}
\vskip .1cm
\caption{\label{fig:gap}
The gap renormalisation due to the depolarisation
level shift. The bare and shifted energy are shown depending on
the state angular momentum, which dependence is given
schematically. The states below $L_F$ are occupied. The states
above are free according to the closed--shell model.}
\end{figure}

Within the closed--shell model the optical gap occurs
between the levels $|L_F\rangle$ and $|L_F+1\rangle$
(see Fig.\ref{fig:gap}) with the value:
\begin{equation}
\displaystyle E_g^{(o)}=\frac{\hbar \omega_o}{2}
\displaystyle \left[ (L_F+1)(L_F+2) -L_F(L_F+1) \right]
= \hbar \omega_o (L_F+1).
\label{SMQWgap1}
\end{equation}
The gap value does depend on the cluster size, decreasing to the
zero as $N$ going to infinity in order to approach the gapless
graphite. For the buckminsterfullerene C$_{60}$ ($R\sim 3.6$~\AA)
the orbital energy quantum is $\hbar \omega_o\sim 0.3$~eV, and
the Fermi momentum is about $4-5$ (the uncertainty is
due to the exact number of $\pi$ electrons is more than 50 for
$L_F=4$ and less than 72 for $L_F=5$). Then the estimation for
the one--electron gap, $1.5-1.8$~eV, is in a reasonable agreement
with the experimental value about $1.8$~eV. We note that the
cluster radius has to be a fitting parameter due to
the 2D approximation lying in the base of the SMQW model.

The gap should increase owing to the zero--point oscillations.
It is because of
the higher level shifts faster. The energy difference between
$L_F$ and $L_F+1$ levels reads as:
\begin{equation}
E_g=\hbar \omega_o
(L_F+1) \left( 1+
2 \frac{(L_F+1)^2}{N}  \kappa \right),
\label{gap2}
\end{equation}
where the parameter $\kappa\simeq 0.36$ is the same as before.
Within the closed--shell approximation,
similarly to what done to get Eq.(\ref{maxls}),
the Fermi momentum follows from the condition:
$N= 2 \sum\limits_{L=0}^{L_F} (2L+1)= 2 (L_F+1)^2$,
because of the number of the occupied states is one half of the
total number of states.
Thus Eq.(\ref{gap2}) becomes extremely simple and contains
no fitting parameters. The gap correction is universal for any
closed--shell
spherical cluster and amounts about 40 \% to the bare value:
\begin{equation}
E_g=E_g^{(o)} ( 1+ \kappa)\simeq 1.36 E_g^{(o)}.
\label{gap3}
\end{equation}

\section{Conclusions}

In the paper we deduced the semiclassical theory of the {\em
depolarization} level shift (LS) in the electronic
spherical--shell system for the fullerene nanoclusters. The LS
materializes by the adiabatic interaction of the charge
carrier with the local field of the fast zero--point oscillations
of the plasmon modes {\em confined} to the cluster spherical
surface. The analytical expression was derived for the
depolarization LS. It is shown that the perturbation theory is
applicable even for the highest (unoccupied) electron level which
has the largest shift. The LS depends on the cluster size as well
as on the angular momentum of the one--electron state. As a
function of the bare one--electron energy all levels shifts, even
for different clusters, collapse onto a straight line
(Fig.\ref{fig:parabola}).
Though,
for the scaling, solely the fixed state with $L^2/N=$const (for
example, the state at the Fermi level) has a physical sense. The
relative energy correction of this fixed state is the same for a
cluster of any size. The more the number of atoms, the higher the
Fermi momentum, while the ratio $L_F^2/N$, contained in the
expression for the depolarization, remains a constant number
about $1/2$. Therefore, the Fermi level correction is independent
of the size. This universal law for the LS which is non--equal
for the states above and below the Fermi level (see
Fig.\ref{fig:gap}) gives a universal rise to the one--electron
gap. As the result the renormalized gap in any closed--shell
spherical cluster is wider by 1.36 times.

\vskip 12pt

{\bf Acknowledgments. } This work was partially supported by
RFBR grants no. 96-15-96348 and 99-02-18170.


\begin{references}

\bibitem{rotkftt}
Rotkin V.V., Suris R.A.,
Sov.- Solid State Physics, vol. {\bf 36} (12), 1899
, 1994.

\bibitem{dress}
Saito R., Dresselhaus G., Dresselhaus M.S.,
Phys.Rev. {\bf B 46} (15), 9906
, 1992.

\bibitem{bertch}
Yabana K., Bertsch G.F.,
Physica Scripta, vol. {\bf 48}, 633
, 1993.

\bibitem{assa}
Murthy G.N., Auerbach A.
Phys.Rev. {\bf B 46} (1), 331
, 1992.

\bibitem{barton}
Barton G., Eberlein C.,
J. Chem.Phys.,vol. {\bf 95}, N 3, 1512-1517, 1991.
V.V.Rotkin, R.A.Suris,
%
%
Fullerenes. Recent Advances in the Chemistry and Physics
of Fullerenes and Related Materials. Vol. III.  Editors:
R.S.Ruoff and K.M.Kadish. ECS Inc.,
Pennington, NJ; PV 96-10, 940
, 1996.


\bibitem{rkfp}
Rotkin V.V., Suris R.A.,
%
Russian Conference of Physics of Semiconductors, Zelenogorsk,
St.Petersburg, vol. 1, p. 68,
26.2--1.3, 1996.
Bechstedt F., Fiedler M., Sham L.J.
Phys.Rev. {\bf B 59} (3), 1857
, 1999.


\bibitem{varshal}
D.A. Varshalovich, A.N. Moskalev, V.K. Khersonskii, Quantum
theory of the angular momentum. Leningrad: Nauka, 1975 (Russian).

\bibitem{bulgak}
Ju N., Bulgac A., Phys.Rev.{\bf B 48} (12), 9071, 1993.

\bibitem{molmat}
V.V.Rotkin, R.A.Suris,
Mol. Materials, v.4, 87-94, 1994.
V.V.Rotkin, R.A.Suris,
Ioffe Institute Prize Winners`97, p.26-33, St.Petersburg, Russia, 1998
(Preprint of Ioffe Institute).


\bibitem{plasm}
Lambin Ph., Lucas A.A., Vigneron J.--P., Phys.Rev. {\bf B 46},
1794, 1992.
%
%
Michalewicz M.T., Das M.P., Sol.State Comm. {\bf 84}, 1121, 1992.
Bulgac A., Ju N., Phys.Rev. {\bf B 46}, 4297, 1992.

\bibitem{hawaii}
Rotkin S.V.,
%
%
Procs of the First International Symposium on
Advanced Luminescent Materials and Quantum Confinement,
Editors: M. Cahay, S. Bandyopadhyay,
D.J. Lockwood, J.P. Leburton, N. Koshida, M. Meyyappan,
and T. Sakamoto.
ECS Inc., Pennington, NJ, PV 99-22,
369
, 1999.

\bibitem{migdal}
A.B. Migdal, Qualitative methods in quantum theory, Moskow:
Nauka, 1975 (Russian).


\bibitem{bz}
No standard 2D Brilluene zone can be defined for
the spherical excitation. The only contraction limit
($R,N,L\to\infty$) allows to relate the $|L,M\rangle$ states to
$| |{\vec k}|,\arg {\vec k}\rangle$ states of the graphene sheet.
Then the maximum value of $k$ equal to $\pi/b\sqrt{3}$.



\end{references}
\end{document}